\documentstyle[11pt,aaspp4]{article}
\newcommand\GRB{GRB~970508}
\newcommand\D{{\cal D}}

\begin{document}

\title{Decelerating Plasmoid Model for Gamma-Ray Burst Afterglows}

\author{James Chiang\altaffilmark{1} \& Charles D. Dermer}

\affil{E. O. Hulburt Center for Space Research, Code 7653,\\ 
       Naval Research Laboratory, Washington, DC 20375-5352}
\altaffiltext{1}{NAS/NRC Resident Research Associate}

\begin{abstract}

The flaring and fading radio, optical, and X-ray afterglows from
\GRB\ are modeled by a highly relativistic plasma sphere which
decelerates by sweeping up ambient gas.  The afterglow emission is
assumed to be synchrotron radiation emitted by nonthermal electrons in
the magnetized plasmoid. The temporal behavior of the delayed emission
is controlled by the evolution of the Doppler factor and by adiabatic
expansion losses of the nonthermal electrons in the plasmoid.  Model
fits to the optical data of \GRB\ are provided, and the relative delay
of the radio peak to the optical peak is found to result from a
decrease in the observed self-absorption frequency as the plasmoid
expands and decelerates.  A variety of afterglow behaviors occurs for
different observing angles and plasmoid parameters. The degree of
collimation inferred from our fit to \GRB\ implies a space density of
GRB sources which exceeds the estimates from scenarios involving
coalescing compact objects. This model can be verified through
observations of superluminal motion in the delayed radio emission.

\end{abstract}


\section{Introduction}

Beppo-SAX observations (Heise 1997) of fading X-ray emission have
resulted in the first identified radio and optical counterparts to
gamma-ray bursts (GRBs).  Five GRBs have been localized with the Wide
Field Camera of the Beppo-SAX experiment to within 3 arc minutes (one
during the science verification phase), with 3 showing X-ray
afterglows.  Optical counterparts have been identified for GRB 970228
and \GRB, with a flaring radio counterpart identified with the latter
GRB (Frail et al.  1997).  Detection of optical Fe~II and Mg~II
absorption lines in the spectrum of the optical counterpart to
\GRB\ (Metzger et al. 1997) demonstrates that at least one GRB is
at cosmological distances with $z \ge 0.835$.  Even though GRB~970111
was the most intense of the five Beppo-SAX bursts, an X-ray afterglow
was not detected from it.  This indicates that very different
behaviors should be expected from different GRBs.

The high energy emission from GRBs is almost certainly radiated by
non-thermal particles as evidenced by their rapid variabilty and
broken power-law spectral shapes (e.g., Fishman \& Meegan 1995). The
spectra of the afterglows also indicate a non-thermal origin.  The
spectrum of the GRB~970228 afterglow has a flux density $F_\nu \propto
\nu^{-1/2}$ extending from the optical to the X-rays (Katz, Piran \& Sari
1997).  The optical spectrum of the \GRB\ afterglow measured at May
10.178 UT (30.6 hours after the initial burst at May~8.904~UT) is also
a power-law with energy index $\alpha = 0.65$ (Djorgovski et
al. 1997).  Furthermore, the optical/X-ray spectral index found by
interpolating the R-band fluxes at the time of the Beppo-SAX NFI X-ray
observation at May~9.1375~UT is consistent with this value.  The radio
spectrum of \GRB\ measured on May~13.96 UT rises from 1.43 to 8.46~GHz
with index $\alpha \approx -1.1$ ($S_\nu \propto \nu^{-\alpha}$; Frail
et al. 1997) which suggests a self-absorbed synchrotron spectrum with
self-absorption frequency near 10~GHz.  Taken together, these data
argue in favor of a nonthermal synchrotron process for the afterglow
emission.

The inferred isotropic total energy of \GRB\ at $z = 0.835$ is
$\approx 10^{52}$ ergs (e.g., Waxman 1997), which exceeds the energy
available through $\nu$-$\bar\nu\rightarrow e^+$-$e^-$ processes in
neutron-star coalescence events (Ruffert et al. 1997) by $\gtrsim 3$
orders of magnitude.  Blast-wave scenarios involving the impulsive
release of energy into a thin, spherically expanding shell also
encounter difficulties in explaining the complex time profiles of GRBs
under the assumption of local spherical symmetry for the blast wave
(Fenimore, Madras \& Nayakshin 1996). Relativistic outflows consisting
of discrete emitting components relieve the energetics problems and
may also ameliorate the difficulties in explaining GRB light curves.

In this {\it Letter}, we consider a model for GRB afterglows where the
emission is produced by nonthermal electrons entrained in collimated,
highly relativistic magnetized plasmoids. The initial flaring of the
emission at frequencies greater than the self-absorbtion frequency is
due to the opening angle of the beaming cone intercepting the
line-of-sight as the plasmoid slows down. At lower frequencies, the
light curve is also affected by the evolution of the self-absorption
frequency.  Subsequent fading of the emission results from bulk
deceleration of the plasmoid and the corresponding decrease of the
Doppler factor with time.  We provide model fits to the optical light
curve of \GRB\ which bracket the delayed radio emission and which are
in accord with the measured X-ray emission. If the model parameters
used to fit \GRB\ are representative of a substantial fraction of
GRBs, then implications for the number of GRB sources and the nature
of the GRB emission mechanism follow.


\section{Dynamics and Synchrotron Emission for a Homogeneous 
Plasma Sphere}

For simplicity we approximate the geometry of the emitting plasma as a
single, homogeneous sphere, although the afterglow emitting region
could have a more complex geometry or be composed of many separate
emission elements moving with a range of speeds.  Let $R_0$ be the
initial comoving radius of the plasmoid at time $t^*_0$ as measured in
the frame of the burst explosion.  If most of the energy content of
the plasmoid is contained in a nonrelativistic thermal component and
the plasmoid expands adiabatically with thermal speed $v_{\rm th}(t) =
v_{\rm th,0} R_0/R(t)$, then $R(t) = [R_0^2 + 2 v_{\rm th,0} R_0
t]^{1/2}$, where $t$ is the time measured in the comoving frame of the
plasmoid.  Conservation of momentum for the plasmoid in the limit of
negligible internal energy loss due to radiation (Katz 1994; see Katz
and Piran 1997 for the opposite limit) implies that $m(x)
\beta(x)\Gamma(x) = m_0 \beta_0\Gamma_0$, where $m(x) = m_0
+\int_{x_0}^x dx^\prime \rho(x^\prime)A(x^\prime)$ is the total mass
at location $x$, $\rho(x)$ is the mass density of ambient gas, and
$A[x(t)]=\pi R^2(t)$ is the cross-sectional area of the plasmoid.
Here $\Gamma(x)$ is the bulk Lorentz factor of the plasmoid at $x$,
and $\Gamma_0 = \Gamma(x_0)$. The relation between the time elements
in the comoving and explosion frame is just $\delta t^* =\Gamma\delta
t$.  These expressions are solved iteratively to determine the speed
of the plasmoid as a function of time.

We assume that nonthermal electrons are described in the comoving
frame by an isotropic angular distribution and a power-law
shock-acceleration spectrum given by $N_e(\gamma_0) \equiv
N_0\gamma_0^{-p}$ for $1\leq\gamma_0\leq \gamma_{\rm max}$, where
$\gamma_0$ is the injection electron Lorentz factor. These electrons
lose energy by adiabatic and synchrotron processes. The adiabatic
energy-loss rate for relativistic electrons is given by
$-\dot\gamma_{\rm adi} \cong \gamma \dot V/(3V) = \gamma\dot
R/R$. When such losses dominate,
\begin{equation}
N_e[\gamma(t)] = N_0\gamma^{-p}[{ R(t)\over R_0}]^{1-p},\qquad 
     1 \leq \gamma \leq \gamma_{\rm max}{ R_0\over R(t)},
\end{equation}
where $N_0 = E_{\rm nt} (2-p)/[m_ec^2 (\gamma^{2-p}_{\rm max}-1)]$ 
and $E_{\rm nt}$ is the energy injected into the nonthermal electrons.

The optically thin synchrotron flux density is given by
\begin{equation}
S_{\rm syn}(\epsilon;\Omega) = m_e c^2{\D}^{3+\alpha}\;
      {N_0 k_{\rm syn}(p) \epsilon_{\rm B}\over d_L^2}\;
       (1+z)^{1-\alpha}\;({\epsilon\over\epsilon_{\rm B}})^{-\alpha}\; ,
       \;1\lesssim{\epsilon (1+z) \over{\D}\epsilon_{\rm B}}
                   \lesssim \gamma_{\rm max}^2,
\label{thin_synchrotron}
\end{equation}
where ${\D} = [\Gamma(1-\beta\mu)]^{-1}$ is the Doppler factor, $d_L$
is the luminosity distance to the source, $\alpha = (p-1)/2$,
$\epsilon = h\nu/(m_e c^2)$, $\epsilon_{\rm B}(t) = B(t)/(4.414\times
10^{13}$ Gauss), and $B(t) = B_0 [R_0/R(t)]^{2q}$ with $q=1$ for
flux-freezing. The angle between the beaming axis and the observer is
$\theta = \cos^{-1}\mu$, and $k_{\rm syn}(p) = (3/2)^\alpha
a(p)\alpha_f^2 c/(2\pi r_e)$ where $\alpha_f$ is the fine structure
constant, $r_e$ is the classical electron radius, and $a(p)$ is a
combination of $\Gamma$-functions given by Blumenthal
\& Gould (1970). The self-absorption frequency is
\begin{equation}
\epsilon_m(t) = \frac{{\D}\epsilon_{\rm B}}{1+z}
                \left[\frac{9}{8\pi}\frac{\sigma_T c(p) N_0}
                {\alpha_f d(p) \epsilon_{\rm B} R^2}\right]^{2/(p+4)}.
\end{equation}
Here $c(p)$ is another combination of $\Gamma$-functions and $d(p)$ is
the optical depth to synchrotron self-absorption through the center of
the plasmoid at the self-absorption frequency; both these functions
are described and tabulated by Gould (1979).  Above and below the
self-absorption frequency, the observed spectrum can be well-described
by a broken power-law:
\begin{eqnarray}
S_{\rm syn}(\epsilon;\Omega) 
     &= & S_0 \left(\frac{\epsilon}{\epsilon_m}\right)^{-\alpha}, \qquad
          \max(\epsilon_{\rm B}, \epsilon_m) \le \epsilon
             \le \epsilon_{\rm B}\gamma_{\rm max}^2\\
     &= & S_0 \left(\frac{\epsilon}{\epsilon_m}\right)^{-5/2}, \qquad
          \epsilon < \epsilon_m,
\end{eqnarray}
where the normalization $S_0$ is given by
Equation~\ref{thin_synchrotron}.  We also calculate the synchrotron
self-Compton component (see Dermer, Sturner, \& Schlickeiser 1997) and
find that it makes negligible contribution to the radio through X-ray
emission of the \GRB\ afterglow.

\section{Model Fits for the \GRB\ Afterglow Light Curves}

The optical afterglow light curve of \GRB\ was found to rise following
the burst event; it then peaked after $\sim 2$~days, and thereafter
decayed with an approximate $t_{\rm obs}^{-1}$ dependence similar to
the decay law of the GRB~970228 optical afterglow.  The most extensive
set of flux measurements for \GRB\ are in the R-band, for which
precise photometric estimates have been made for a subset of these
observations (Djorgovski et al. 1997).  The R-band data is shown in
Figure~1a along with our approximate fits for two sets of burst and
plasmoid parameters.

In the context of our decelerating plasmoid model, the shape of the
afterglow light curve can be understood analytically by considering
the simpler case where the ambient density and plasmoid
cross-sectional area are constants.  The peak in the optical light
curve occurs when the bulk Lorentz factor has decreased to a value of
$\Gamma_p \simeq 1/\theta$.  Note that if the observing angle
$\theta\lesssim 1/\Gamma_0$, then there will be no initial rise in the
light curve of the non-self-absorbed emission as observed for the
initial optical afterglow of GRB~970228.  The ratio
$F_{\nu,p}/F_{\nu,0}$ of the peak flux and the flux at early time is
proportional to $(\D_p/\D_0)^{3+\alpha}$
(cf. Equation~\ref{thin_synchrotron}), where $\D_p$ and $\D_0$ are the
Doppler factors at the time of the peak flux and immediately following
the burst, respectively.  For $\Gamma^2 \gg 1$ and $\theta \ll 1$, the
Doppler factor is $\D \cong 2\Gamma/(1 + \Gamma^2\theta^2)$ so that
the ratio of peak to intial Doppler factor is $\D_p/\D_0 \cong
2\Gamma_0\theta/(1 + \Gamma_0^2\theta^2)$.  Using this result and the
flux ratio found from the R-band light curve, $F_{\nu,p}/F_{\nu,0}
\simeq 4$, we solve for $\theta$ and find $\theta
\simeq 2.6/\Gamma_0$.

The time delay of the peak, $\tau_{\rm obs}$, constrains the
cross-sectional area $A_0$ of the plasmoid and the density of the
ambient medium.  From momentum balance, we have
$\Gamma_p/\Gamma_0\cong (1+A_0\rho c \tau^*/m_0)^{-1}$.  This gives
\begin{equation}
A_0 \cong \frac{\Gamma_0\theta - 1}{\Gamma_0} 
      \frac{E_0\theta^2}{2 \rho c^3 \tau_{\rm obs}},
\end{equation}
where we have used the approximation $\tau_{\rm obs} \cong
\tau^*\theta^2/2$.  The expansion of the plasmoid has the effect of
increasing the required value of $\theta$ and reducing the required
initial cross-sectional area of the plasmoid relative to the
non-expanding case. All non-self-absorbed emission below the
high-energy cutoffs should rise and fall with the same characteristic
behavior in this model.

In producing the fits shown in Figure~1a, we have fixed the total
plasmoid energy $E_0 =10^{50}$~ergs (which is dominated by the kinetic
energy of the thermal plasma, i.e., $E_0 = \Gamma_0 m_0 c^2$), set $p
= 2.3$ in accord with the measured optical spectral index of $\alpha =
0.65$ (Djorgovski et al.  1997), and considered bulk initial Lorentz
factors for the plasmoid in the range $\Gamma_0 = 10^2$--$10^3$.  We
have also considered initial thermal expansion velocities in the range
$v_{\rm th} = 10^7$--$3\times 10^9$~cm~s$^{-1}$ and ambient densities
of $\rho = 0.1$--$1$~cm$^{-3}\times m_p$.  The initial energy of
nonthermal electrons is determined by an injection efficiency defined
by $\eta \equiv E_{\rm nt}/E_0$. The complete set of parameters for
the two fits is given in Table~1.  We note that the magnetic field
strengths of the plasmoid in both cases are well below the
equipartition values, which ensures that the self-absorption frequency
is not too high.  Furthermore, we find that these relatively low
values for the magnetic field are also consistent with our assumption
that the electron spectrum evolves mainly through adiabatic cooling,
since synchrotron cooling is only important at frequencies
\begin{equation}
\nu_{\rm obs}~({\rm Hz}) \gtrsim \frac{3\times 10^{16}}
                                      {(\D /100) (B/0.01~{\rm G})^3
                                       (t_{\rm obs}/10^6~{\rm s})^2}.
\label{synch_cooling_frequencies}
\end{equation}

Just below the self-absorption frequency, expansion of the plasmoid
plays an important role in detemining the location of the peak of the
light curve.  For the afterglow of \GRB, the peak of the radio
emission is delayed by a few days relative to the peak of the optical
light curve, and this is consistent with the inference that the
self-absorption frequency is in this range.  As the plasmoid expands,
it becomes non-self-absorbed at increasingly lower frequencies.  This
produces an additional enhancement in the observed flux at these
frequencies superposed on the flaring of the emission due to the
deceleration of the plasmoid, which also causes the observed
self-absorption frequency to decrease with time.  The 8.4--8.46~GHz
radio data (Frail et al.~1997) shown in Figure~1b is bracketed by our
model results which fit the optical light curve in Figure~1a.  The
parameters in Table~1 cover a wide range of values and have not been
fine-tuned to fit the radio data precisely.  However, the model fits
pass through the X-ray data point (Piro et al.~1997) shown in
Figure~1c without any additional adjustment.  Our model predicts that
the X-ray and optical light curves would rise and fall together,
except that the X-ray flux can also decline as a result of synchrotron
cooling (cf. Eq.~[\ref{synch_cooling_frequencies}]) or, depending on
the initial value of $\gamma_{\rm max}$, through either adiabatic or
synchrotron losses.

\section{Discussion}

We have proposed a model for GRB afterglows where nonthermal electrons
emit synchrotron radiation in relativistic bulk plasma outflows.
Other models for GRB afterglows also consider synchrotron processes
(e.g., Wijers, Rees, \& M\'esz\'aros 1997; Waxman 1997; Katz \& Piran
1997; Panaitescu et al. 1997; Tavani 1997), but these studies treat a
blast wave scenario, whereas we consider highly collimated
outflows. This significantly reduces the energetics requirements and
the need for the emitting region to be radiatively efficient. Indeed,
we do not find that the radiating plasmoid in our model for \GRB\
slows to nonrelativistic speeds.

The small beaming angles and large Lorentz factors derived here have
important implications for GRB models under the assumption that the
prompt gamma-ray emission from a typical GRB has a beaming angle
$\theta_{\rm GRB} \simeq 1/\Gamma_0$.  This assumption is reasonable
because there is no evidence from GRB light curves for spreading of
the pulse widths during the main portion of the burst, which would be
expected if the emitting plasma was significantly decelerating during
the gamma-ray active phase.  We find that the inferred total gamma-ray
energy for \GRB\ is reduced by a factor $\sim 10^3$ from the isotropic
value if the gamma-ray beaming pattern is $\propto \D_0^{3+\alpha}$,
noting also that the gamma-ray emission is being viewed at an angle
$\theta \approx 2.6/\Gamma_0$ from the beaming axis.  Furthermore,
because the gamma-rays are beamed into a fraction $\sim
1/(2\Gamma_0^2)$ of the full sky, a much larger source rate than the
isotropic rate is required.  This rate is too large to be accounted
for by the rate of coalescing compact objects (see review by Dermer \&
Weiler 1995).  If this model is correct, it therefore favors other GRB
senarios such as naked collapse events of white dwarfs (Dar et
al. 1992), birth events of highly magnetized neutron stars (Usov
1992), or failed Type 1b supernova models (Woosley 1993).  The large
Lorentz factor found for the radio, optical and X-ray observations of
\GRB\ does not follow the power-law scaling (Paczy\'nski \& Rhoads
1993; M\'esz\'aros \& Rees 1997) of the Lorentz factor with frequency
invoked by Rhoads (1997) to estimate the optical transient detection
rate.  It also contradicts the inference by Rhoads (1997) that the
material which produces the afterglow emission is only mildly
relativistic, and thus removes a central motivation for the hypernova
model of Paczy\'nski (1997) insofar as highly collimated burst
emission models are not ruled out.

If \GRB\ is typical of a large fraction of GRBs and if the gamma-ray
beaming pattern is indeed governed by the Doppler factor $\D_0$ at
early times, then we can make definite predictions for the likelihood
of detecting optical afterglow emission from such bursts.  Figure~2
shows the afterglow light curves for our two sets of model parameters
at different observing angles.  For GRBs viewed at smaller angles than
the inferred observing angle for \GRB\, the afterglow optical flux is
greatly enhanced, though the probability of detecting these bright
afterglows is correspondingly smaller.  For example, we find that the
peak optical afterglow emission from one out of every $\sim 25$ GRBs
will reach R $\simeq 13.3 + 5\log(z/0.8)$, given that the peak R-band
magnitude of \GRB\ was 19.6 (Castro-Tirado et al. 1997). Optical
transients not associated with GRBs (i.e., when $\theta \gg
1/\Gamma_0$) reach a much fainter peak magnitude. This effect is
important for optical transient detection estimates and implies much
lower rates than estimated by Rhoads (1997). As noted previously, the
afterglow light curve (for the non-self-absorbed emission) should fall
monotonically for observing angles $\theta < 1/\Gamma_0$. This can
explain the optical afterglow light curve of GRB~920228 which appears
to decline monotonically.  The plasmoid producing the \GRB\ optical
afterglow between 3 and 10 days after the burst travels 30--100 pc.
If it enters a lower density medium over this distance, its optical
flux will decline much more slowly and could account for the
discrepancy between the model fit and data shown in Figure~1a.  This
effect might also explain the flattening of the GRB~970228 R-band
light curve after $\sim 1$ week (Galama et al. 1997a).

The burst afterglow model considered here is similar to nonthermal
synchrotron models for the radio-through-optical continua of blazars,
though with larger Lorentz factors.  Hence, we expect that certain
blazar characteristics should also be seen in GRB afterglows, such as
superluminal motion and high polarization. In particular, we predict
that radio and optical superluminal motion should be measured.  From
the parameters used to fit \GRB\, we expect a motion of $\sim 0.3$
milliarcseconds per month.  Detection of superluminal motion of this
magnitude in afterglow emissions of GRBs would require highly
relativistic emission regions, in accord with the model proposed here.

\acknowledgements
This work was performed while J.C. held a National Research
Council-NRL Research Associateship.  C.D. acknowledges support by the
Office of Naval Research.



\begin{figure}[h]
\epsfysize=6.25in
\centerline{\epsfbox{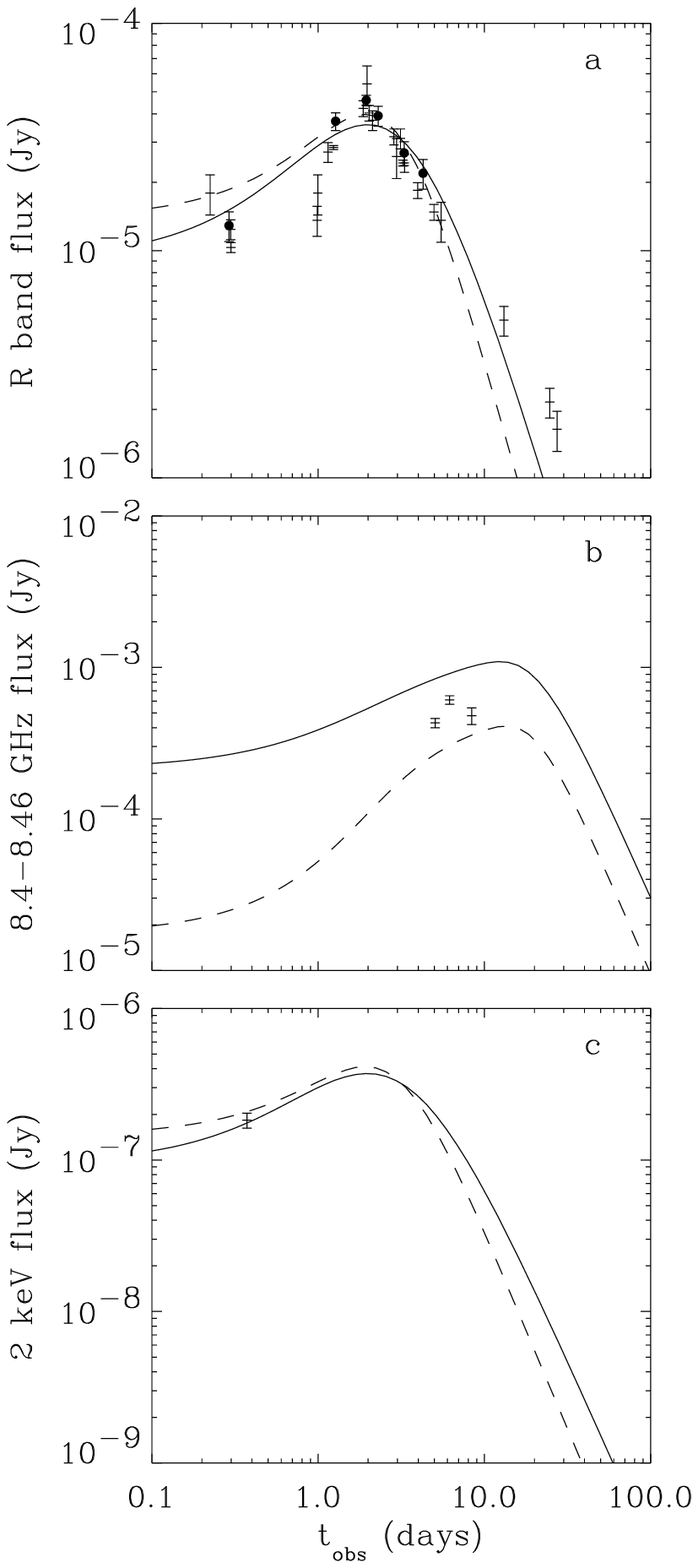}}
\vspace{0.5in}
\caption{Optical, radio and X-ray light curves for the afterglow 
of \protect{\GRB}.  The R-band data from Djorgovski et al (1997) are
plotted as filled circles and the remaining R-band data [from Galama
et al. (1997b), Castro-Tirado et al. (1997), Schaefer et al. (1997),
Groot et al. (1997), Garcia et al. (1997), Chevalier \& Ilovaisky
(1997), Kopylov et al. (1997a \& b), Fruchter et al. (1997), and
Metzger et al. (1997)] are scaled to the measurement of Mignoli et
al. (1997) of R = 19.78 at May 10.85 which has been photometrically
calibrated by Djorgovski et al..  The 8.4--8.64 GHz radio data are
from Frail et al. (1997), and the 2~keV data point is from Piro et
al. (1997).  The solid curves are for the Model~1 parameters and the
dashed curves are for Model~2.  Note that the model X-ray light curves
may also be affected by a high energy cut-off in the electron
distribution.}
\end{figure}

\begin{figure}[h]
\epsfysize=6.25in
\centerline{\epsfbox{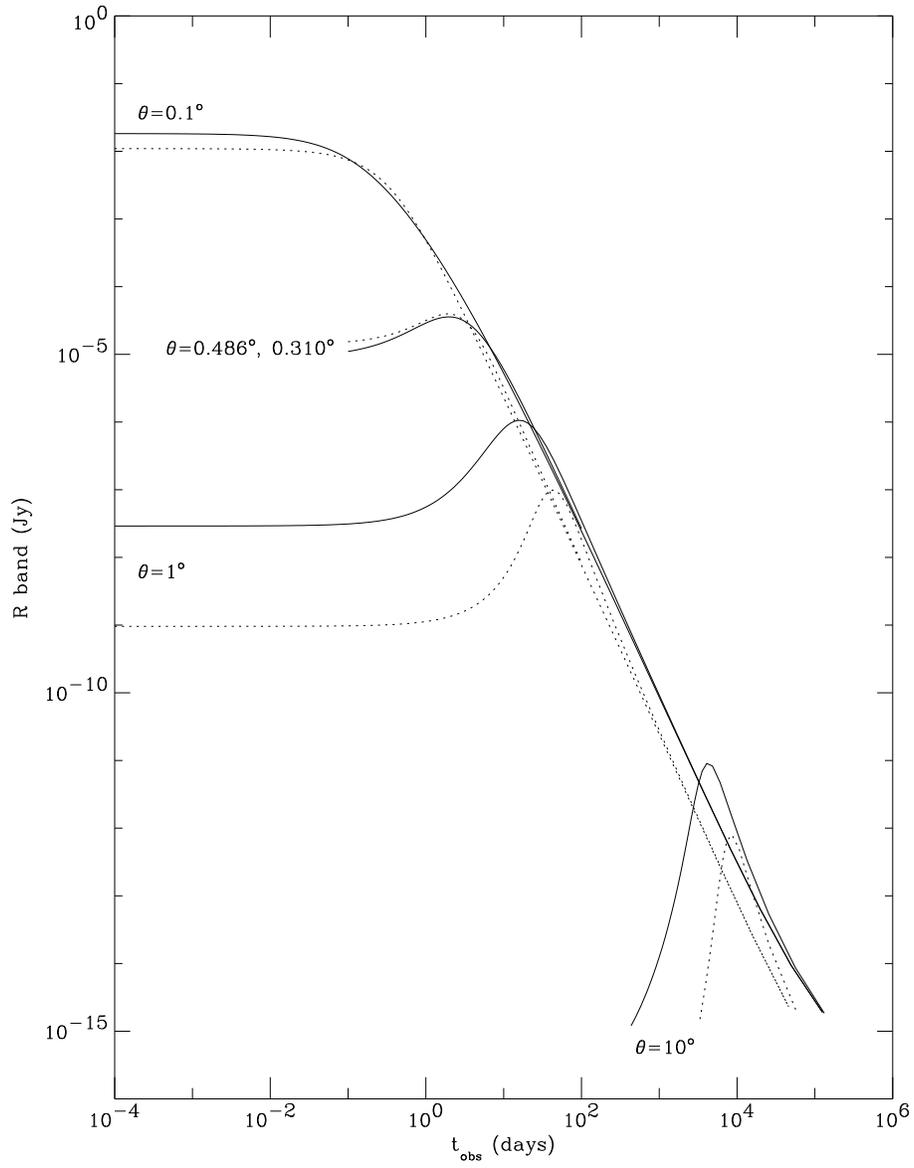}}
\vspace{0.5in}
\caption{R-band light curves at various viewing angles for the two 
sets of model parameters, Model~1 (solid curve) and Model~2 (dotted).}
\end{figure}


\begin{table}[h]
\small\centering
\caption{Decelerating Plasmoid Parameters for the Afterglow Lightcurves 
of \protect{\GRB}}
\bigskip
\begin{tabular}{cccccccc}
\tableline\tableline
Model & $\Gamma_0$ & $\theta\,(^\circ)$ & $A_0$ (cm$^2$) &
   $\rho/m_p$ (cm$^{-3}$) & $v_{\rm th}$ (cm~s$^{-1}$) & $B_0$ (G) & $\eta$ \\
\tableline
1 & 300 & 0.486 & $2.5\times10^{31}$ & 0.1 & $3\times10^7$ & $10^{-2}$ & 0.02\\
2 & 500 & 0.310 & $4.0\times10^{29}$ & 1.0 & $1\times10^7$ & $10^{-3}$ & 0.4 \\
\tableline
\end{tabular}
\label{beta_values}
\end{table}

\end{document}